\documentclass[prb,aps,superscriptaddress,twocolumn,dvips]{revtex4}
\usepackage{feynmp}
\usepackage{amsmath}
\usepackage{amssymb}
\usepackage{epsfig}
\usepackage{graphicx}
\usepackage{subfigure}
\usepackage{hyperref}
\usepackage{bbm}
\usepackage{color}
\usepackage{longtable}
\usepackage{booktabs}
\usepackage{multirow}
\usepackage{array}
\newcommand{\PreserveBackslash}[1]{\let\temp=\\#1\let\\=\temp}
\newcolumntype{C}[1]{>{\PreserveBackslash\centering}p{#1}}
\newcolumntype{R}[1]{>{\PreserveBackslash\raggedleft}p{#1}}
\newcolumntype{L}[1]{>{\PreserveBackslash\raggedright}p{#1}}
\newcommand{\tabincell}[2]{\begin{tabular}{@{}#1@{}}#2\end{tabular}}

\newcommand{\Rmnum}[1]{\expandafter\@slowromancap\romannumeral #1@}
\allowdisplaybreaks

\begin{document}
\author{Hai-Xiao Xiao}
\affiliation{Department of Physics, Nanjing University, Nanjing
210093, P. R. China}

\author{Jing-Rong Wang}
\affiliation{Anhui Province Key Laboratory of Condensed Matter
Physics at Extreme Conditions, High Magnetic Field Laboratory of the
Chinese Academy of Science, Hefei, Anhui 230031, P. R. China}

\author {Guo-Zhu Liu}
\altaffiliation{gzliu@ustc.edu.cn} \affiliation{Department of Modern
Physics, University of Science and Technology of China, Hefei, Anhui
230026, P. R. China}

\author{Hong-Shi Zong}
\altaffiliation{zonghs@nju.edu.cn} \affiliation{Department of
Physics, Nanjing University, Nanjing 210093, China}
\affiliation{Joint Center for Particle, Nuclear Physics and
Cosmology, Nanjing, Jiangsu 210093, P. R. China} \affiliation{State
Key Laboratory of Theoretical Physics, Institute of Theoretical
Physics, CAS, Beijing 100190, P. R. China}

\title{Robustness of the semimetal state of Na$_{3}$Bi and Cd$_{3}$As$_{2}$
against Coulomb interaction}

\begin{abstract}
We study the excitonic semimetal-insulator quantum phase transition
in three-dimensional Dirac semimetal in which the fermion dispersion
is strongly anisotropic. After solving the Dyson-Schwinger equation
for the excitonic gap, we obtain a global phase diagram in the plane
spanned by the parameter for Coulomb interaction strength and the
parameter for fermion velocity anisotropy. We find that excitonic
gap generation is promoted as the interaction becomes stronger, but
is suppressed if the anisotropy increases. Applying our results to
two realistic three-dimensional Dirac semimetals Na$_{3}$Bi and
Cd$_{3}$As$_{2}$, we establish that their exact zero-temperature
ground state is gapless semimetal, rather than excitonic insulator.
Moreover, these two materials are far from the excitonic quantum
critical point, thus there should not be any observable evidence for
excitonic insulating behavior. This conclusion is in general
agreement with the existing experiments of Na$_{3}$Bi and
Cd$_{3}$As$_{2}$.
\end{abstract}

\maketitle

\section{INTRODUCTION}

There has been increasing research interests in the physical
properties of three-dimensional Dirac semimetal (3D DSM) that
contains massless Dirac fermions at low energies \cite{Vafek14,
Wehling14, Armitage17}. Such DSM state could emerge at the quantum
critical point (QCP) between normal insulator and 3D topological
insulator. Interestingly, 3D DSM has been observed in
TiBiSe$_{2-x}$S$_{x}$ \cite{XuSuYang11, Sato11} and
Bi$_{2-x}$In$_{x}$Se$_{3}$ \cite{Brahlek12, Wu13} by fine tuning the
doping level. Theoretical studies \cite{WangZhiJun12, Wangzhijun13}
predicted that a crystal-symmetry protected stable 3D DSM might be
realized in such materials as A$_{3}$Bi (A=Na, K, Rb) and
Cd$_{3}$As$_{2}$. Recent angle-resolved photoemission spectroscopy
(ARPES) and quantum transport measurements reported evidences for the
existence of 3D DSM state in Na$_{3}$Bi and Cd$_{3}$As$_{2}$
\cite{LiuZK14A, Neupane14, LiuZK14B, Borisenko14, HeLP14}.

Similar to 2D DSM \cite{Vafek14, Wehling14, CastoNeto09, Kotov12}
and other semimetals \cite{Armitage17, Yan17, Hasan17, XuFang11,
Fang12, YangNogasa14}, 3D DSM contains a number of discrete
band-touching points, which means that the density of states (DOS)
vanishes at the Fermi level. As a result, the Coulomb interaction
between massless fermions is poorly screened and remains long-ranged
\cite{Goswami11, Hosur12, Hofmann15, Throckmorton15, Sekine14,
Gonzalez14, Gonzalez15, Braguta16, Braguta17}. Extensive theoretical
studies on 2D DSM, with graphene being a prominent example, have
revealed that a sufficiently strong long-range Coulomb interaction
can induce excitonic-type pairing and as such opens a dynamical gap
at the Fermi level \cite{Kotov12, CastroNetoPhysics09,
Khveshchenko01, Gorbar02, Khveshchenko04, Liu09, Khveshchenko09,
Gamayun10, Sabio10, Liu11, WangLiu11A, WangLiu11B, WangLiu12,
Popovici13, WangLiu14, Carrington16, Xue17,Sharma17, Xiao17, Gamayun09,
WangJianhui11, Katanin16, Vafek08, Gonzalez10, Gonzalez12, Drut09A,
Drut09B, Drut09C, Armour10, Armour11, Buividovich12, Ulybyshev13,
Smith14, Tupitsyn17, Juan12, Kotikov16}. The particle-hole
condensate breaks the chiral symmetry of the system
\cite{Khveshchenko01, Gorbar02}, which is a condensed-matter
realization of the non-perturbative phenomenon of dynamical chiral
symmetry breaking that plays an essential role in hadron physics
\cite{Roberts94}. Once a finite dynamical gap is generated, the
semimetal state becomes unstable and the system is converted into an
insulator \cite{Kotov12, CastroNetoPhysics09}. A nature question is
whether a similar excitonic insulating transition also occurs in a
3D DSM.

The possible semimetal-insulator transition in 3D DSM has been
studied by several groups \cite{Sekine14, Gonzalez14, Gonzalez15,
Braguta16, Braguta17}. In Na$_{3}$Bi and Cd$_{3}$As$_{2}$, the
$z$-component of the fermion velocity is considerably smaller than
the other two components within the $x$-$y$ plane \cite{LiuZK14A,
Neupane14, LiuZK14B}. Additionally, the magnitude of fermion
velocity in these two materials is quite small. This implies that
the Coulomb interaction may play a significant role at low energies.
After performing Monte Carlo simulations, Braguta \emph{et al.}
\cite{Braguta16, Braguta17} claimed that both Na$_{3}$Bi and
Cd$_{3}$As$_{2}$ lie deep in the excitonic insulating phase. This
conclusion is somewhat surprising, because experiments did not find
any evidence for the insulating behavior in these two materials
\cite{LiuZK14A, Neupane14, LiuZK14B, Borisenko14}. It is necessary
to examine whether the gapless semimetal state is robust against the
long-range Coulomb interaction in Na$_{3}$Bi, Cd$_{3}$As$_{2}$, and
other candidate 3D DSM materials.

In order to determine the true ground state of 3D DSM, we need to
calculate the critical value of the Coulomb interaction strength
that separates the semimetallic and insulating phases. For
Na$_{3}$Bi and Cd$_{3}$As$_{2}$, the energy dispersion of 3D Dirac
fermions can be written as
\begin{eqnarray}
E = \pm\sqrt{v_{\parallel}^{2}k_{\parallel}^{2}+v_{z}^{2}k_{z}^{2}},
\end{eqnarray}
where $k_{\parallel}^{2} = k_{x}^{2} + k_{y}^{2}$. Here,
$v_{\parallel}$ is the component of fermion velocity within the
basal $x$-$y$ plane, and $v_{z}$ is the component along the
$z$-direction. The effective strength of the Coulomb interaction is
represented by the parameter \cite{Kotov12}
\begin{eqnarray}
\alpha &=& \frac{e^{2}}{v_{\parallel}\epsilon_{0}\epsilon_{r}},
\end{eqnarray}
where $e$ is the electron charge, $\epsilon_{0}$ the vacuum
dielectric constant, and $\epsilon_{r}$ the relative dielectric
constant. The value of $\epsilon_{r}$ is strongly material
dependent. It is known \cite{Gorbar02, Khveshchenko04, Liu09,
Khveshchenko09, Gamayun10, WangLiu12, Drut09A, Drut09B, Drut09C}
that an excitonic gap is generated only when $\alpha$ is larger than
certain critical value $\alpha_c$. If $\alpha > \alpha_c$, the
system has an insulating ground state, which can be detected by
probing the transport properties at ultra low temperatures
\cite{Geim11, Geim12}. If $\alpha$ is slightly smaller than
$\alpha_c$, the exact zero-temperature ground state is semimetal.
However, since the system is close to the excitonic insulating QCP,
the quantum fluctuation of excitonic order parameter could be
important at small distances, which may still have observable
effects \cite{Hirata17}. If $\alpha \ll \alpha_c$, the system is
deep in the semimetal phase, and does not exhibit any observable
effects of insulating behavior.

In this paper, we calculate the value of $\alpha_c$ in 3D DSM by
using the non-perturbative Dyson-Schwinger (DS) equation method
\cite{Khveshchenko01, Gorbar02, Khveshchenko04, Liu09,
Khveshchenko09, Gamayun10, Sabio10, Liu11, WangLiu11A, WangLiu11B,
WangLiu12, Popovici13, WangLiu14, Carrington16, Sharma17, Xiao17,
Appelquist88, Maris04, Feng06, Feng12A, Feng12B, LiJianFeng13,
Janssen16, WangLiuZhang17, Gusynin16, Fischer18}. In some 3D DSMs,
such as Na$_{3}$Bi and Cd$_{3}$As$_{2}$, the fermion dispersion is
strongly anisotropic, and the $z$-component of fermion velocity is
much smaller than that of the $x$-$y$ plane, namely $v_{z} \ll
v_{\parallel}$. We need to define a velocity ratio and study how the
ratio affects $\alpha_c$. A commonly used definition
\cite{Braguta16, Braguta17, Sekine14} is
\begin{eqnarray}
\eta = \frac{v_{z}}{v_{\parallel}}.
\end{eqnarray}
After solving the DS gap equation numerically, we obtain a global
phase diagram of 3D DSM in the parameter space spanned by $\alpha$
and $\eta$. It is found that $\alpha_{c}$ exhibits a non-monotonic
dependence on the velocity anisotropy, analogous to what happens in
2D DSM \cite{Xiao17}. We demonstrate that such non-monotonic
dependence results from the improper definitions of $\alpha$ and
$\eta$ utilized in previous works. We then introduce a physically
more appropriate definition for these parameters, and show that the
velocity anisotropy is indeed detrimental to the formation of
excitonic pairing. As a direct application of our result, we
establish that Na$_{3}$Bi and Cd$_{3}$As$_{2}$ are actually both
deep in the semimetal phase, which is well consistent recent
experiments \cite{LiuZK14A, Neupane14, LiuZK14B, Borisenko14}.

The complete set of DS equations cannot be exactly solved without
employing certain truncation scheme. Here, we first solve the DS
equation for dynamical gap by entirely ignoring both fermion
velocity renormalization and wave-function renormalization. The
critical value $\alpha_c$ obtained by employing this truncation is
larger than the physical value of $\alpha$ in Na$_{3}$Bi and
Cd$_{3}$As$_{2}$. We then move to examine the influence of
higher-order corrections. In particular, we include the dynamical
screening of Coulomb interaction, the fermion velocity
renormalization, the wave-function renormalization, and also the
vertex correction into the DS equations. Our calculations reveal
that, although $\alpha_c$ is more or less altered by higher-order
corrections, the conclusion that Na$_{3}$Bi and Cd$_{3}$As$_{2}$ are
both deep in the semimetal phase remains intact.

The rest of the paper is structured as follows. In
Sec.~\ref{Sec:Model}, we present the DS equation for the dynamical
gap by employing a number of different approximations. In
Sec.~\ref{Sec:NumResults}, we solve the DS equations and discuss the
physical implication of our results. In this section, we also
introduce a more suitable definition for $\alpha$ and $\eta$, which
allows us to examine the impact of Coulomb interaction and velocity
anisotropy separately. The influence of higher-order corrections is
analyzed in Sec.~\ref{Sec:HighOrderCorrection}. A brief summary of
our results is given in Sec.~\ref{Sec:Summary}.

\section{Model and gap equation\label{Sec:Model}}

The free Hamiltonian of 3D Dirac fermions is
\begin{eqnarray}
H_{0} = \int d^{3}\mathbf{r}\bar{\Psi}_{a}(\mathbf{r})
\left(v_{x}\gamma_{1}\nabla_{x}+v_{y}\gamma_{2}\nabla_{y} +
v_{z}\gamma_{3}\nabla_{z}\right)\Psi_{a}(\mathbf{r}), \nonumber \\
\end{eqnarray}
where $\Psi_{a}$ is a four-component spinor and $\bar{\Psi}_{a} =
\Psi^{\dag}\gamma_{0}$. The index $a = 1, 2, .., N$ with $N$ being
the fermion flavor. For Na$_{3}$Bi and Cd$_{3}$As$_{2}$, the
physical flavor is $N=2$, corresponding to the two Dirac cones in
the Brillouin zone \cite{Braguta16, Braguta17}. We will consider a
general large flavor $N$ in order to perform $1/N$ expansion. The
gamma matrices $\gamma_{\mu}$, with $\mu=0,1,2,3$, are defined in
the standard way, satisfying the Clifford algebra
$\{\gamma_{\mu},\gamma_{\nu}\} = 2\delta_{\mu\nu}$. For 3D DSM
materials Na$_{3}$Bi and Cd$_{3}$As$_{2}$, $v_{x} = v_{y}$, but
$v_{z}$ takes an obviously different value. In the following, we
assume that $v_{x} = v_{y} = v_{\parallel}$. The long-range Coulomb
interaction between Dirac fermions is described by
\begin{eqnarray}
H_{ee} &=& \frac{1}{4\pi}\int d^{3}\mathbf{r}
d^{3}\mathbf{r}'\bar{\Psi}_{a}(\mathbf{r}) \gamma_{0}
\Psi_{a}(\mathbf{r})\frac{e^2}{\epsilon_{0} \epsilon_{r}
\left|\mathbf{r}-\mathbf{r}'\right|}\nonumber \\
&&\times \bar{\Psi}_{a}(\mathbf{r}')\gamma_{0}\Psi_{a}(\mathbf{r}').
\end{eqnarray}
The total Hamiltonian $H_0 + H_{ee}$ preserves a continuous chiral
symmetry $\Psi_{a}\rightarrow e^{i\theta\gamma_{5}}\Psi_{a}$, where
$\theta$ is an arbitrary constant and $\gamma_{5} =
\gamma_{0}\gamma_{1}\gamma_{2}\gamma_{3}$, which will be broken once
a finite excitonic gap $m \propto \langle
\bar{\Psi}_{a}\Psi_{a}\rangle$ is dynamically generated by the
Coulomb interaction.

The bare fermion propagator has the form
\begin{eqnarray}
G_{0}(\varepsilon,\mathbf{p}) = \frac{1}{\varepsilon\gamma_{0} +
v_{\parallel}(\gamma_{1}p_{x}+\gamma_{2}p_{y})+v_{z}\gamma_{3}p_{z}}.
\label{Eq:BareFermionPropagator}
\end{eqnarray}
The dressed Coulomb interaction can be expressed as
\begin{eqnarray}
V(\Omega,q_{\parallel},q_{z}) = \frac{1}{V_{0}(\mathbf{q}) +
\Pi(\Omega,q_{\parallel},q_{z})},\label{Eq:CoulombExp}
\end{eqnarray}
where the bare Coulomb interaction is
\begin{eqnarray}
V_{0}(\mathbf{q}) = \frac{\mathbf{q}^2}{4 \pi\alpha v_{\parallel}},
\end{eqnarray}
and $\Pi(\Omega,q_{\parallel},q_{z})$ is the polarization
function.

Due to the Coulomb interaction, the free fermion propagator is
strongly renormalized to become
\begin{eqnarray}
G\left(\varepsilon,\mathbf{p}\right) =
\frac{1}{G_{0}^{-1}\left(\varepsilon,\mathbf{p}\right) -
\Sigma(\varepsilon,\mathbf{p})}, \label{Eq:DysonEq}
\end{eqnarray}
where $G(\varepsilon,\mathbf{p})$ is the full fermion propagator.
The fermion self-energy $\Sigma(\varepsilon,\mathbf{p})$ is given by
\begin{eqnarray}
\Sigma(\varepsilon,\mathbf{p}) &=& \int\frac{d\omega}{2\pi}
\frac{d^3\mathbf{k}}{(2\pi)^2}
\Gamma(\varepsilon,\mathbf{p};\omega,\mathbf{k})\gamma_{0}
G(\omega,\mathbf{k})\gamma_{0}\nonumber \\
&\times& V(\varepsilon-\omega,\mathbf{p}-\mathbf{k}),
\label{Eq:Selfenergy}
\end{eqnarray}
where $\Gamma(\varepsilon,\mathbf{p};\omega,\mathbf{k})$ is the
vertex function. Generically, the self-energy can be formally
expressed as
\begin{eqnarray}
\Sigma(\varepsilon,\mathbf{p}) &=& \left(1-A_{0}\right)
\gamma_{0}\varepsilon + \left(1-A_{1}\right)\left(\gamma_{1}p_{x} +
\gamma_{2}p_{y}\right)v_{\parallel} \nonumber \\
&& +\left(1-A_{2}\right)\gamma_{3}p_zv_{z} + m,
\end{eqnarray}
which then leads to
\begin{eqnarray}
G(\varepsilon,\mathbf{p}) = \frac{1}{A_{0}\gamma_{0}\varepsilon +
A_{1}v_{\parallel}(\gamma_{1}p_{x} + \gamma_{2}p_{y}) +
A_{2}v_{z}\gamma_{3}p_z + m}. \label{Eq:Fullfermionpropagator}
\end{eqnarray}
Here, $A_{0,1,2}\equiv A_{0,1,2}(\varepsilon,p_{\parallel},p_{z})$
are three wave function renormalization factors and $m\equiv
m(\varepsilon,p_{\parallel},p_{z})$ denotes the dynamical excitonic
gap. The Landau damping of fermions is embodied in the function
$A_0$, whereas the renormalization of fermion velocities can be
obtained from $A_{1}$ and $A_{2}$. The model can be treated by means
of $1/N$ expansion \cite{Appelquist88, Maris04}.

We will first solve the DS equations by retaining the leading-order
of $1/N$ expansion, and then examine the influence of higher-order
corrections. To the leading-order, one can set $A_{0,1,2} \equiv 1$.
Accordingly, the vertex function can be taken as $\Gamma \equiv
1$, as required by the Ward identity. Combining the above
several equations, we derive the following DS gap equation
\begin{widetext}
\begin{eqnarray}
m(\varepsilon,p_{\parallel},p_{z}) &=& \int
\frac{d\omega}{2\pi} \frac{d^{3}\mathbf{k}}{(2\pi)^{3}}
\frac{m(\omega,k_{\parallel},k_{z})}{\omega^{2} +
v_{\parallel}^{2}k_{\parallel}^{2} + v_{z}^{2}k_{z}^{2} +
m^{2}(\omega,k_{\parallel},k_{z})} V(\varepsilon-\omega,
(\mathbf{p}-\mathbf{k})_{\parallel},p_{z}-k_{z}).
\label{Eq:GapExp}
\end{eqnarray}
If this equation has only vanishing solution, namely $m \equiv 0$,
the zero-temperature ground state is strictly gapless and the
semimetal phase is robust against Coulomb interaction. If a nonzero
solution for $m$ is obtained, a finite fermion gap is dynamically
generated, leading to excitonic insulating transition. To solve the
gap equation, we still need to know the detailed expression of
dressed Coulomb interaction. As shown in Appendix A, to the leading
order of $1/N$ expansion, the polarization function can be well
approximated by
\begin{eqnarray}
\Pi(\Omega,q_{\parallel},q_{z}) =
\frac{N\left(v_{\parallel}^{2}q_{\parallel}^{2} +
v_{z}^{2}q_{z}^{2}\right)}{6\pi^2v_{\parallel}^{2}v_{z}}
\ln\left(\frac{\left(v_{\parallel}^{2}v_{z}\right)^{1/3}\Lambda +
\sqrt{\Omega^{2}+v_{\parallel}^{2}q_{\parallel}^{2} + v_{z}^{2}
q_{z}^{2}}}{\sqrt{\Omega^{2}+v_{\parallel}^{2}q_{\parallel}^{2} +
v_{z}^{2}q_{z}^{2}}}\right), \label{Eq:PolaExp}
\end{eqnarray}
where $\Lambda$ is the momentum cutoff. The derivation of
$\Pi(\Omega,q_{\parallel},q_{z})$ is given in
Appendix~\ref{Sec:AppendixPola}. Making use of
Eq.~(\ref{Eq:CoulombExp}), Eq.~({\ref{Eq:GapExp}}), and
Eq.~(\ref{Eq:PolaExp}), we obtain the following gap equation
\begin{eqnarray}
m(\varepsilon,p_{\parallel},p_{z}) &=& \int \frac{d\omega}{2\pi}
\frac{d^{3}\mathbf{k}}{(2\pi)^{3}}
\frac{m(\omega,k_{\parallel},k_{z})}{\omega^{2} +
k_{\parallel}^{2}+\eta^{2}k_{z}^{2} +
m^{2}(\omega,k_{\parallel},k_{z})} \frac{1}{\frac{|\bf{q}|^{2}}{4
\pi\alpha} + \frac{N\left(q_{\parallel}^{2} +
\eta^{2}q_{z}^{2}\right)}{6\pi^2\eta} \ln\left(\frac{\eta^{1/3} +
\sqrt{\Omega^{2}+q_{\parallel}^{2} + \eta^{2}
q_{z}^{2}}}{\sqrt{\Omega^{2}+q_{\parallel}^{2} +
\eta^{2}q_{z}^{2}}}\right)}, \label{Eq:GapetaExp}
\end{eqnarray}
where $\Omega = \varepsilon-\omega$ and $\mathbf{q} =
\mathbf{p}-\mathbf{k}$. To derive this equation, we have made the
following re-scaling transformations:
\begin{eqnarray}
\frac{p_{\parallel}}{\Lambda}\rightarrow p_{\parallel},\,\,
\frac{k_{\parallel}}{\Lambda}\rightarrow k_{\parallel},\,\,
\frac{q_{\parallel}}{\Lambda}\rightarrow q_{\parallel},\,\,
\frac{p_{z}}{\Lambda} \rightarrow p_{z},\,\, \frac{k_{z}}{\Lambda}
\rightarrow k_{z},\,\, \frac{q_{z}}{\Lambda} \rightarrow q_{z},\,\,
\frac{\varepsilon}{v_{\parallel}\Lambda}\rightarrow\varepsilon,\;\,
\frac{\omega}{v_{\parallel}\Lambda}\rightarrow\omega,\,\,
\frac{\Omega}{v_{\parallel}\Lambda}\rightarrow\Omega,\,\,
\frac{m}{v_{\parallel}\Lambda}\rightarrow m.
\end{eqnarray}
\end{widetext}

The dynamical gap is a function of three variables, namely
$\varepsilon$, $p_{\parallel}$, and $p_{z}$. Given the non-linear
nature of Eq.~(\ref{Eq:GapetaExp}), it is extremely difficult to
solve the equation numerically without making further
approximations. Here, we will adopt two widely used approximations.
The first one is the instantaneous approximation, which neglects the
energy dependence of Coulomb interaction
\begin{eqnarray}
m(\varepsilon,p_{\parallel},p_{z}) &\rightarrow&
m(p_{\parallel},p_{z}), \\
V(\Omega,\mathbf{q}) &\rightarrow& V(0,\mathbf{q}).
\end{eqnarray}
Accordingly, the gap function becomes energy independent, i.e.,
\begin{eqnarray}
m(\varepsilon,p_{\parallel},p_{z}) &\rightarrow&
m(p_{\parallel},p_{z}).
\end{eqnarray}

Under this approximation, it is straightforward to integrate over
$\omega$, which yields a simplified gap equation:
\begin{eqnarray}
m(p_{\parallel},p_{z}) &=& \frac{1}{2} \int
\frac{d^{3}\mathbf{k}}{(2\pi)^{3}}
\frac{m(k_{\parallel},k_{z})}{\sqrt{k_{\parallel}^2
+ \eta^{2}k_{z}^{2} + m^{2}(k_{\parallel},k_{z})}}\nonumber \\
&\times&\frac{1}{\frac{\mathbf{q}^{2}}{4\pi\alpha} +
\frac{N\left(q_{\parallel}^{2}+\eta^{2}q_{z}^{2}\right)}{6\pi^2\eta}
\ln\left(\frac{\eta^{1/3} + \sqrt{q_{\parallel}^{2} +
\eta^{2}q_{z}^{2}}}{\sqrt{q_{\parallel}^{2} +
\eta^{2}q_{z}^{2}}}\right)}.\nonumber \\
\label{Eq:GapEquationInstantaneous}
\end{eqnarray}
The dynamical screening is ignored in this equation.

To incorporate the dynamical screening effect, Khveshchenko
\cite{Khveshchenko09} proposed a different approximation, which
assumes that the energy dependence of dynamical screening is assumed
to be equivalent to the momenta dependence. Under the Khveshchenko
approximation, the gap equation takes the form
\begin{eqnarray}
m(p_{\parallel},p_{z}) &=& \frac{1}{2}\int
\frac{d^{3}\mathbf{k}}{(2\pi)^{3}}
\frac{m(k_{\parallel},k_{z})}{\sqrt{k_{\parallel}^2 +
\eta^{2}k_{z}^{2} + m^{2}(k_{\parallel},k_{z})}} \nonumber \\
&\times&\frac{1}{\frac{\mathbf{q}^{2}}{4\pi\alpha} + \frac{N
\left(q_{\parallel}^{2}+\eta^{2}q_{z}^{2}\right)}{6\pi^2\eta}
\ln\left(\frac{\eta^{1/3}+\sqrt{2(q_{\parallel}^{2} +
\eta^{2}q_{z}^{2})}}{\sqrt{2(q_{\parallel}^{2} +
\eta^{2}q_{z}^{2})}}\right)}. \nonumber \\
\label{Eq:GapEquationKhveshchenko}
\end{eqnarray}

The gap equations (\ref{Eq:GapEquationInstantaneous}) and
(\ref{Eq:GapEquationKhveshchenko}) can be numerically solved by
using the iteration method. There are two tuning parameters: flavor
$N$ and interaction strength $\alpha$. Theoretically, for an
excitonic gap to be dynamically generated, $N$ should be smaller
than $N_c$ and $\alpha$ should be larger than $\alpha_c$. Once $N_c$
is greater than the physical value, here $N=2$, one can fix $N=2$,
and determine the critical value $\alpha_c$ by varying $\eta$. At
other cases, it is necessary to calculate $N_c$ accordingly for
different $\eta$.

The above two gap equations are derived by retaining the
leading-order contribution of the $1/N$ expansion. The functions
$A_{0,1,2}$ are simply set to unity. This amounts to entirely
neglect the wave-function renormalization and also the fermion
velocity renormalization. According to the extensive DS equation
studies carried out in the context of 2D DSM \cite{WangLiu12,
Popovici13, Carrington16, Kotikov16, Fischer18}, including these
effects might change the value of $\alpha_c$. It is also interesting
to examine how these effects alter the leading-order result of
$\alpha_c$ in 3D DSM.

We now incorporate higher-order contributions to the DS equations.
After substituting Eq.~(\ref{Eq:Selfenergy}) and
(\ref{Eq:Fullfermionpropagator}) into Eq.~(\ref{Eq:DysonEq}), we
obtain four self-consistently coupled equations for
$A_{0,1,2}(\varepsilon,\mathbf{p})$ and $m(\varepsilon,\mathbf{p})$:
\begin{widetext}
\begin{eqnarray}
A_{0}(\varepsilon,\mathbf{p})&=& 1-\frac{1}{\epsilon}\int
\frac{d\omega}{2\pi} \frac{d^{3}\mathbf{k}}{(2\pi)^{3}}
\Gamma(\varepsilon,\mathbf{p};\omega,\mathbf{k})
\frac{A_{0}(\omega,\mathbf{k})\omega}{A_{0}^{2}(\omega,\mathbf{k})
\omega^{2} + A_{1}^{2}(\omega,\mathbf{k})k_{\parallel}^{2} +
A_{2}^{2}(\omega,\mathbf{k}) \eta^{2}k_{z}^{2} +
m^{2}(\omega,\mathbf{k})}V(\Omega,{\mathbf{q}}),
\label{Eq:refullDSEqRGV1} \\
A_{1}(\varepsilon,\mathbf{p}) &=& 1+\frac{1}{p_{\parallel}^{2}}\int
\frac{d\omega}{2\pi} \frac{d^{3}\mathbf{k}}{(2\pi)^{3}}
\Gamma(\varepsilon,\mathbf{p};\omega,\mathbf{k})
\frac{A_{1}(\omega,\mathbf{k})\vec{p}_{\parallel}\cdot
\vec{k}_{\parallel}}{A_{0}^{2}(\omega,\mathbf{k})
\omega^{2}+A_{1}^{2}(\omega,\mathbf{k})k_{\parallel}^{2} +
A_{2}^{2}(\omega,\mathbf{k})\eta^{2}k_{z}^{2} +
m^{2}(\omega,\mathbf{k})} V(\Omega,{\mathbf{q}}),
\label{Eq:refullDSEqRGV2} \\
A_{2}(\varepsilon,\mathbf{p})&=&1+\frac{1}{p_{z}}\int
\frac{d\omega}{2\pi} \frac{d^{3}\mathbf{k}}{(2\pi)^{3}}
\Gamma(\varepsilon,\mathbf{p};\omega,\mathbf{k})
\frac{A_{2}(\omega,\mathbf{k})k_{z}}{A_{0}^{2}(\omega,\mathbf{k})\omega^{2}
+ A_{1}^{2}(\omega,\mathbf{k})k_{\parallel}^{2} +
A_{2}^{2}(\omega,\mathbf{k}) \eta^{2}k_{z}^{2} +
m^{2}(\omega,\mathbf{k})}V(\Omega,{\mathbf{q}}),
\label{Eq:refullDSEqRGV3}\\
m(\varepsilon,\mathbf{p})&=&\int\frac{d\omega}{2\pi}
\frac{d^{3}\mathbf{k}}{(2\pi)^{3}}
\Gamma(\varepsilon,\mathbf{p};\omega,\mathbf{k})
\frac{m(\omega,\mathbf{k})}{A_{0}^{2}(\omega,\mathbf{k})\omega^{2} +
A_{1}^{2}(\omega,\mathbf{k})k_{\parallel}^{2}+A_{2}^{2}(\omega,\mathbf{k})
\eta^{2}k_{z}^{2}+m^{2}(\omega,\mathbf{k})}V(\Omega,{\mathbf{q}}),
\label{Eq:refullDSEqRGV4}
\end{eqnarray}
where $\Omega=\epsilon-\omega$, and $\mathbf{q} =
\mathbf{p}-\mathbf{k}$. To determine the impact of fermion velocity
renormalization, we temporarily ignore the energy dependence of the
dynamical gap, which leads to
\begin{eqnarray}
A_{0}(\varepsilon,\mathbf{p}) = 1, \quad
\Gamma(\varepsilon,\mathbf{p};\omega,\mathbf{k}) = 1.
\label{Eq:VRGInstanCondition}
\end{eqnarray}
Now the above coupled equations can be simplified to
\begin{eqnarray}
A_{1}(p_{\parallel},p_{z}) &=& 1+\frac{1}{p_{\parallel}^{2}}
\frac{1}{2}\int \frac{d^{3}\mathbf{k}}{(2\pi)^{3}}
\frac{A_{1}(k_{\parallel},k_{z})\vec{p}_{\parallel}\cdot
\vec{k}_{\parallel}}{\sqrt{A_{1}^{2}(k_{\parallel},k_{z})
k_{\parallel}^{2} + A_{2}^{2}(k_{\parallel},k_{z})\eta^{2}k_{z}^{2}
+ m^{2}(k_{\parallel},k_{z})}}V(\mathbf{q}), \label{Eq:DSEqRGV1}
\\
A_{2}(p_{\parallel},p_{z}) &=& 1+\frac{1}{p_{z}}\frac{1}{2}\int
\frac{d^{3}\mathbf{k}}{(2\pi)^{3}}\frac{A_{2}(k_{\parallel},k_{z})
k_{z}}{\sqrt{A_{1}^{2}(k_{\parallel},k_{z}) k_{\parallel}^{2} +
A_{2}^{2}(k_{\parallel},k_{z})\eta^{2}k_{z}^{2} +
m^{2}(k_{\parallel},k_{z})}}V(\mathbf{q}), \label{Eq:DSEqRGV2}
\\
m(p_{\parallel},p_{z}) &=& \frac{1}{2}\int \frac{d^{3}
\mathbf{k}}{(2\pi)^{3}}
\frac{m(k_{\parallel},k_{z})}{\sqrt{A_{1}^{2}(k_{\parallel},k_{z})
k_{\parallel}^{2}+A_{2}^{2}(k_{\parallel},k_{z}) \eta^{2}k_{z}^{2} +
m^{2}(k_{\parallel},k_{z})}}V(\mathbf{q}). \label{Eq:DSEqRGV3}
\end{eqnarray}
In these equations, the renormalization of fermion velocity is
encoded in $A_1(p_{\parallel},p_{z})$ and
$A_2(p_{\parallel},p_{z})$, and the Coulomb interaction function is
written as
\begin{eqnarray}
V(\mathbf{q}) = \frac{1}{\frac{\mathbf{q}^{2}}{4\pi\alpha} + \frac{N
\left(q_{\parallel}^{2}+\eta^{2}q_{z}^{2}\right)}{6\pi^2\eta}
\ln\left(\frac{\eta^{1/3}+\sqrt{2(q_{\parallel}^{2} +
\eta^{2}q_{z}^{2})}}{\sqrt{2(q_{\parallel}^{2} +
\eta^{2}q_{z}^{2})}}\right)}.
\end{eqnarray}

We then consider the impact of fermion damping. For this purpose,
the energy dependence of Coulomb interaction should be explicitly
included. For simplicity, we only study the isotropic limit, which
amounts to take $\eta = 1.0$, $v_{||}=v_{z}$, and $A_{1}=A_{2}$. The
coupled DS equations are given by
\begin{eqnarray}
A_{0}(\varepsilon,\mathbf{p}) &=& 1-\frac{1}{\epsilon}\int
\frac{d\omega}{2\pi}\frac{d^{3}\mathbf{k}}{(2\pi)^{3}}
\Gamma(\varepsilon,\mathbf{p};\omega,\mathbf{k})
\frac{A_{0}(\omega,\mathbf{k})\omega}{A_{0}^{2}(\omega,\mathbf{k})\omega^{2}
+ A_{1}^{2}(\omega,\mathbf{k})v_{\parallel}^{2}\mathbf{k}^{2} +
m^{2}(\omega,\mathbf{k})}
V(\Omega,{\mathbf{q}}), \label{Eq:DSESIsotropicA}\\
A_{1}(\varepsilon,\mathbf{p}) &=& 1+\frac{1}{\mathbf{p}^{2}}\int
\frac{d\omega}{2\pi}\frac{d^{3}\mathbf{k}}{(2\pi)^{3}}
\Gamma(\varepsilon,\mathbf{p};\omega,\mathbf{k})
\frac{A_{1}(\omega,\mathbf{k})\mathbf{p}\cdot
\mathbf{k}}{A_{0}^{2}(\omega,\mathbf{k})\omega^{2} +
A_{1}^{2}(\omega,\mathbf{k})v_{\parallel}^{2}\mathbf{k}^{2} +
m^{2}(\omega,\mathbf{k})}V(\Omega,{\mathbf{q}}),
\label{Eq:DSESIsotropicB}\\
m(\varepsilon,\mathbf{p}) &=& \int\frac{d\omega}{2\pi}
\frac{d^{3}\mathbf{k}}{(2\pi)^{3}}
\Gamma(\varepsilon,\mathbf{p};\omega,\mathbf{k})
\frac{m(\omega,\mathbf{k})}{A_{0}^{2}(\omega,\mathbf{k})\omega^{2} +
A_{1}^{2}(\omega,\mathbf{k})v_{\parallel}^{2}\mathbf{k}^{2} +
m^{2}(\omega,\mathbf{k})}V(\Omega,{\mathbf{q}}).
\label{Eq:DSESIsotropicC}
\end{eqnarray}
\end{widetext}
Following Ref.~\cite{Maris04}, we assume that the vertex function
takes the form
\begin{eqnarray}
\Gamma(\varepsilon,\mathbf{p};\omega,\mathbf{k}) =
\frac{1}{2}\left[A_{0}(\varepsilon,\mathbf{p}) +
A_{0}(\omega,\mathbf{k})\right].
\end{eqnarray}
This vertex function is widely in the studies of dynamical chiral
symmetry breaking in QED$_3$ \cite{Maris04} and 2D DSM
\cite{WangLiu12, Fischer18}. The Coulomb interaction function is
\begin{eqnarray}
V(\Omega,\mathbf{q}) = \frac{1}{\frac{\mathbf{q}^{2}}{4 \pi
\alpha}+\frac{N\mathbf{q}^{2}}{6\pi^{2}\eta} \ln \left(\frac{1 +
\sqrt{\Omega^{2}+\mathbf{q}^{2}}}{\sqrt{\Omega^{2} +
\mathbf{q}^{2}}}\right)}.
\end{eqnarray}

All the above DS equations can be numerically solved. The solutions
will be analyzed in the next section.

\section{NUMERICAL RESULTS\label{Sec:NumResults}}

In this section, we present the numerical solutions of the DS
equations obtained under various approximations. As $\alpha$ grows
from a very small value, the excitonic gap is always zero. The gap
develops a nonzero value continuously as $\alpha$ exceeds a critical
value $\alpha_c$, which is identified as the QCP of excitonic
insulating transition. By solving the gap equation at different
values of $\eta$, one can determine how $\alpha_c$ depends on
$\eta$. Moreover, we will introduce a different definition of
$\alpha$ and $\eta$.

\subsection{Instantaneous approximation}

\begin{figure}
\includegraphics[width=0.36\textwidth]{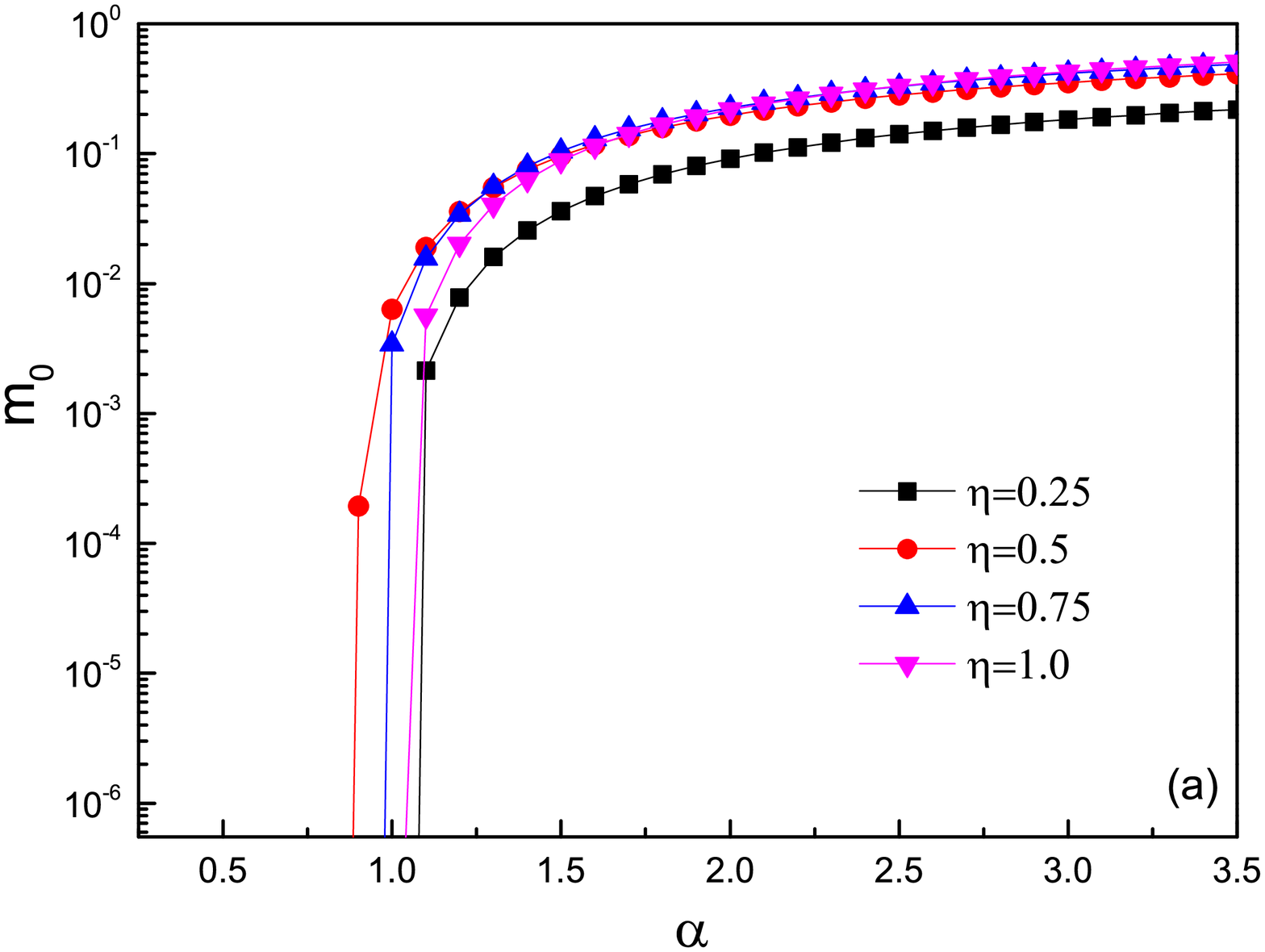}
\includegraphics[width=0.36\textwidth]{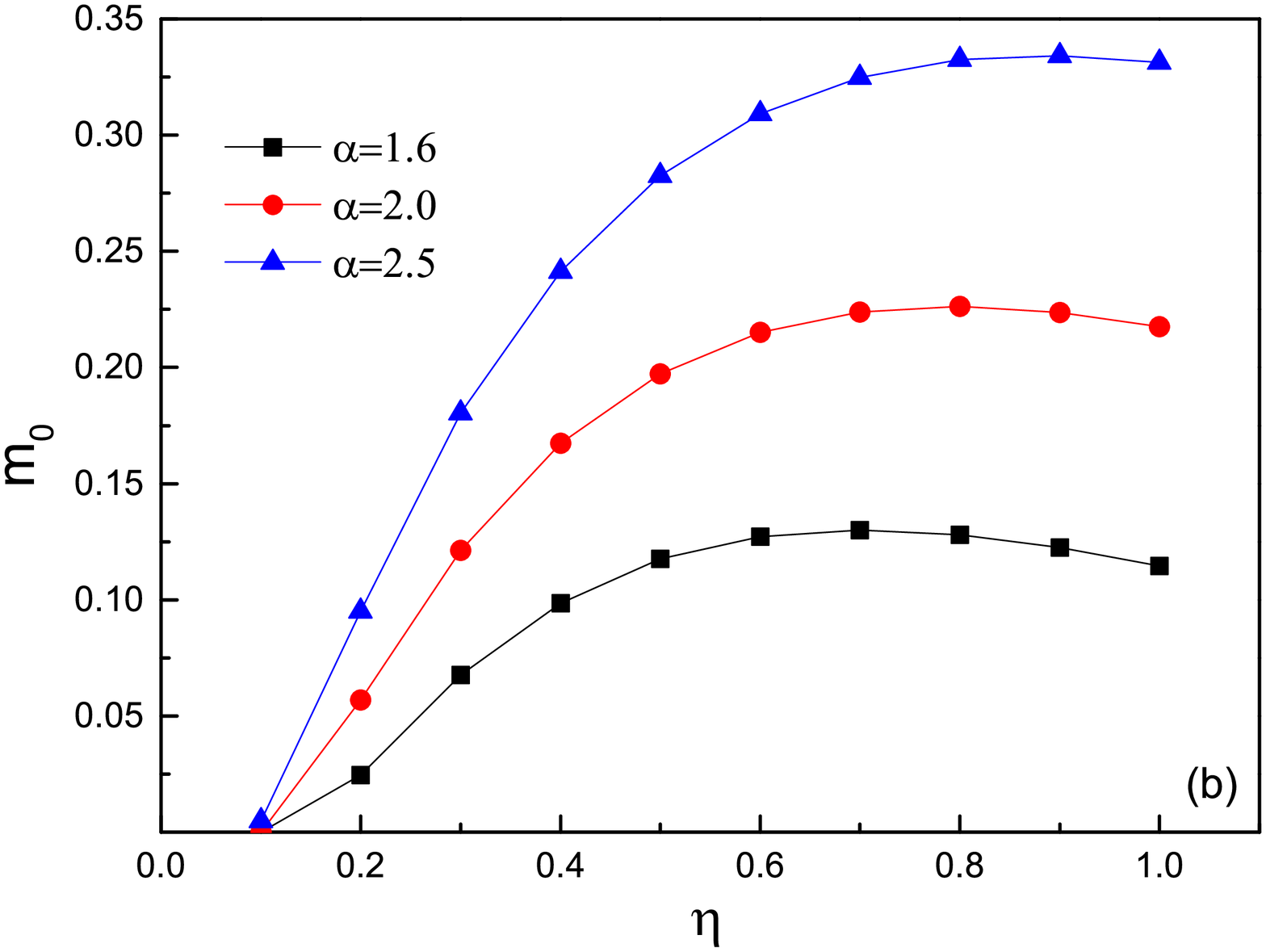}
\caption{(a) The $\alpha$-dependence of $m_{0}$ at different values
of $\eta$; (b) The $\eta$-dependence of $m_{0}$ at different values
of $\alpha$. Results are obtained under the instantaneous
approximation.}\label{Fig:Gap0Instan}
\end{figure}

From the solutions of Eq.~(\ref{Eq:GapEquationInstantaneous}), we
get the zero-energy excitonic gap $m_0$ as a function of $\alpha$
and $\eta$. Though $N_c$ is not accurately determined here, it is
easy to infer that $N_c > 2$, because the gap would always be zero
if $N_c < 2$. In Fig.~\ref{Fig:Gap0Instan}(a), we present the
$\alpha$ dependence of zero-energy gap $m_0$ at several fixed values
of $\eta$. We can see that, once $\alpha$ exceeds a critical value
$\alpha_{c}$, a finite excitonic gap is dynamically generated. The
gap is an monotonously increasing function of $\alpha$. In
Fig.~\ref{Fig:Gap0Instan}(b), we show the $\eta$ dependence of $m_0$
by choosing three different representative values of $\alpha$. From
Fig.~\ref{Fig:Gap0Instan}(b), we observe that, the gap first
increases with the decreasing of anisotropy in the case of strong
anisotropy, but decreases with smaller anisotropy once $\eta$ is
greater than some threshold $\eta_c$. For any given $\alpha$, the
excitonic gap takes its maximal value at $\eta_m$, which depends on
the specific value of $\alpha$. Such non-monotonic $\eta$-dependence
of the gap is caused by the competition between the increase of
Coulomb interaction strength and the increase of velocity
anisotropy. A more detailed explanation will be given in
Sec.~\ref{Numerical:C}.

Based on our numerical results, it is easy to plot a phase diagram
on the $\alpha$-$\eta$ space, as shown in
Fig.~\ref{Fig:PhaseDiagramKhveshchenko}(a). In the isotropic limit
with $\eta = 1$, the critical interaction strength is roughly
$\alpha_{c} \approx 1.1$, which is much smaller than the value
$\alpha_{c} = 1.71$ obtained previously in \cite{Braguta16}, but is
close to the subsequently updated result $\alpha_{c}\approx 1.14$
\cite{Braguta17}.

As an application of our results, we now determine whether the 3D
DSMs Na$_{3}$Bi and Cd$_{3}$As$_{2}$ lie in the semimetal or
excitonic insulating phase. In Table~\ref{Table:FermionVelocities}
and Table~\ref{Table:DielectrincConstant}, we list the concrete
values of the fermion velocities and the relative dielectric
constants in Na$_{3}$Bi and Cd$_{3}$As$_{2}$, respectively. The
physical value of $\alpha$ can be easily estimated from these data.
In previous works \cite{Braguta16, Braguta17}, it was claimed that
$\alpha \approx 7$ in Na$_{3}$Bi and $\alpha \approx 1.8$ in
Cd$_{3}$As$_{2}$. Their calculations did not properly include the
influence of the dielectric constant $\varepsilon_r$. Once
$\varepsilon_r$ is taken into account, the magnitude of $\alpha$
will be substantially reduced. Using the data given in Tables
\ref{Table:FermionVelocities} and \ref{Table:DielectrincConstant},
we find that $\alpha \approx 1.1$ in Na$_{3}$Bi and $\alpha \approx
0.06$ in Cd$_{3}$As$_{2}$. Moreover, it is easy to deduce that $\eta
\approx 0.1$ in Na$_{3}$Bi and $\eta \approx 0.25$ in
Cd$_{3}$As$_{2}$. According to the results presented in
Fig.~\ref{Fig:PhaseDiagramKhveshchenko}(a), $\alpha_{c} \approx 2.1$
for $\eta = 0.1$ and $\alpha_{c} \approx 1.1$ for $\eta = 0.25$.

\begin{table}[htbp]
\caption{Fermion velocities in Na$_{3}$Bi and Cd$_{3}$As$_{2}$
\label{Table:FermionVelocities}}
\begin{center}
\begin{tabular}{|C{1.8cm}|C{1.8cm}|C{1.8cm}|C{1.8cm}|}
\hline\hline Material & $v_{\parallel}\,\mathrm{(m/s)}$ &
$v_{z}\,\mathrm{(m/s)}$ & Reference
\\
\hline Na$_{3}$Bi & $3.74\times10^{5}$ & $2.89\times10^{4}$ &
\cite{LiuZK14A}
\\
\hline Cd$_{3}$As$_{2}$ & \tabincell{c}{$1.5\times10^6$ \\
$1.29\times10^{6}$} & \tabincell{c}{order $10^5$ \\
$3.27\times10^{5}$} & \tabincell{c}{\cite{Neupane14} \\
\cite{LiuZK14B}}
\\
\hline\hline
\end{tabular}
\end{center}
\end{table}
\begin{table}[htbp]
\vspace{-0.5cm} \caption{Relative dielectric constant in Na$_{3}$Bi
and Cd$_{3}$As$_{2}$ \label{Table:DielectrincConstant}}
\vspace{-0.3cm}
\begin{center}
\begin{tabular}{|C{2.4cm}|C{2.4cm}|C{2.4cm}|}
\hline\hline Material & $\epsilon_{r}$ & Reference
\\
\hline Na$_{3}$Bi & $5.9$ & \cite{Mehrdad16}
\\
\hline Cd$_{3}$As$_{2}$ & \tabincell{c}{$20-40$ \\ 30 } &
\tabincell{c}{\cite{Throckmorton15, Zivitz74, JayGerin77} \\
\cite{Adriano17} }
\\
\hline\hline
\end{tabular}
\end{center}
\end{table}

From the above analysis, we immediately deduce that the effective
Coulomb interaction in Na$_{3}$Bi and Cd$_{3}$As$_{2}$ is too weak
to generate an excitonic gap, and that the exact zero-temperature
ground state of these materials is semimetal, rather than excitonic
insulator. Moreover, both Na$_{3}$Bi and Cd$_{3}$As$_{2}$ lie deep
in the gapless semimetallic phase, as shown in
Fig.~\ref{Fig:PhaseDiagramKhveshchenko}(a). There is no detectable
signature of excitonic insulating behavior in these two materials.

\subsection{Khveshchenko approximation}

\begin{figure}
\center
\includegraphics[width=0.363\textwidth]{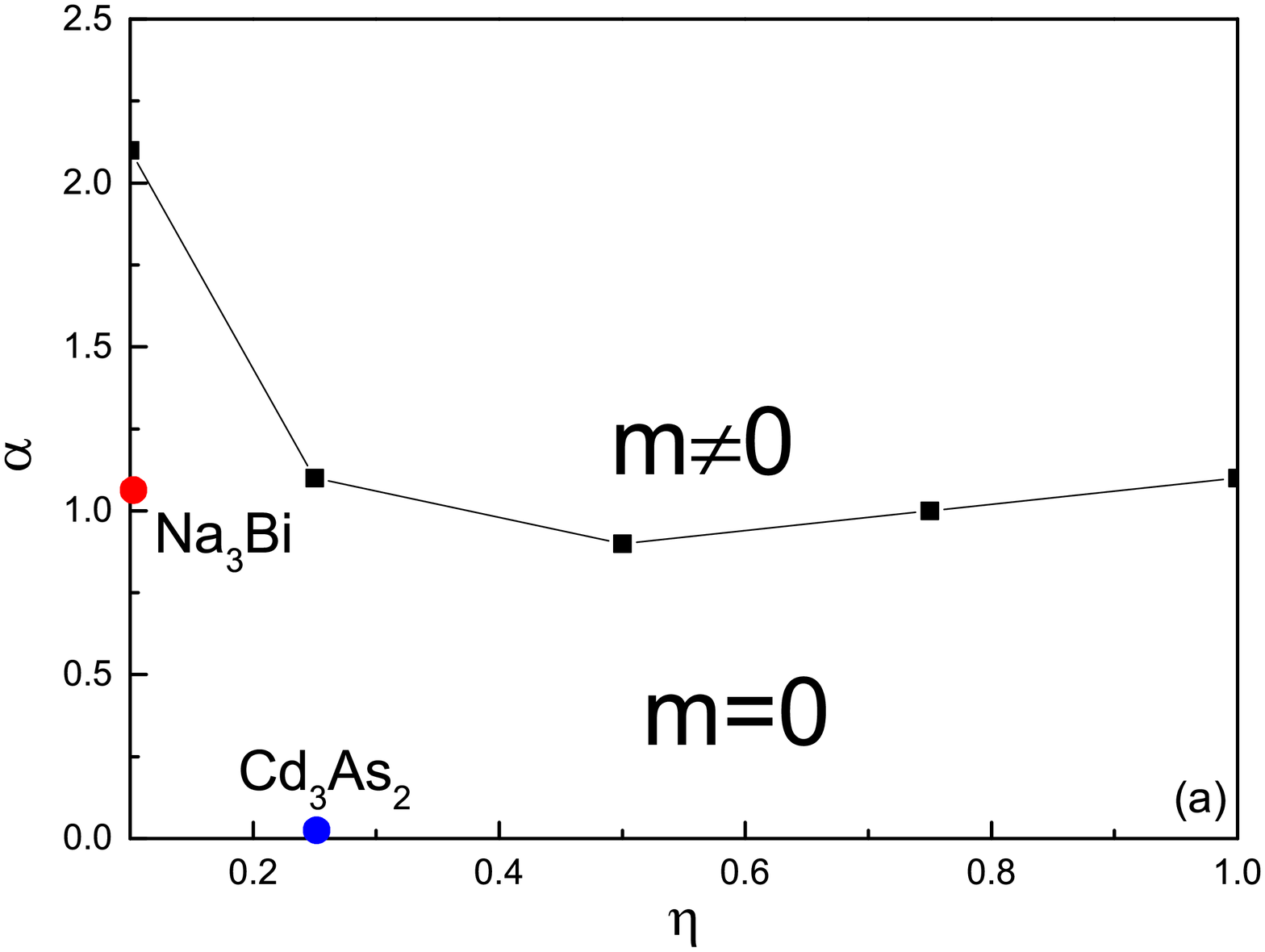}
\includegraphics[width=0.36\textwidth]{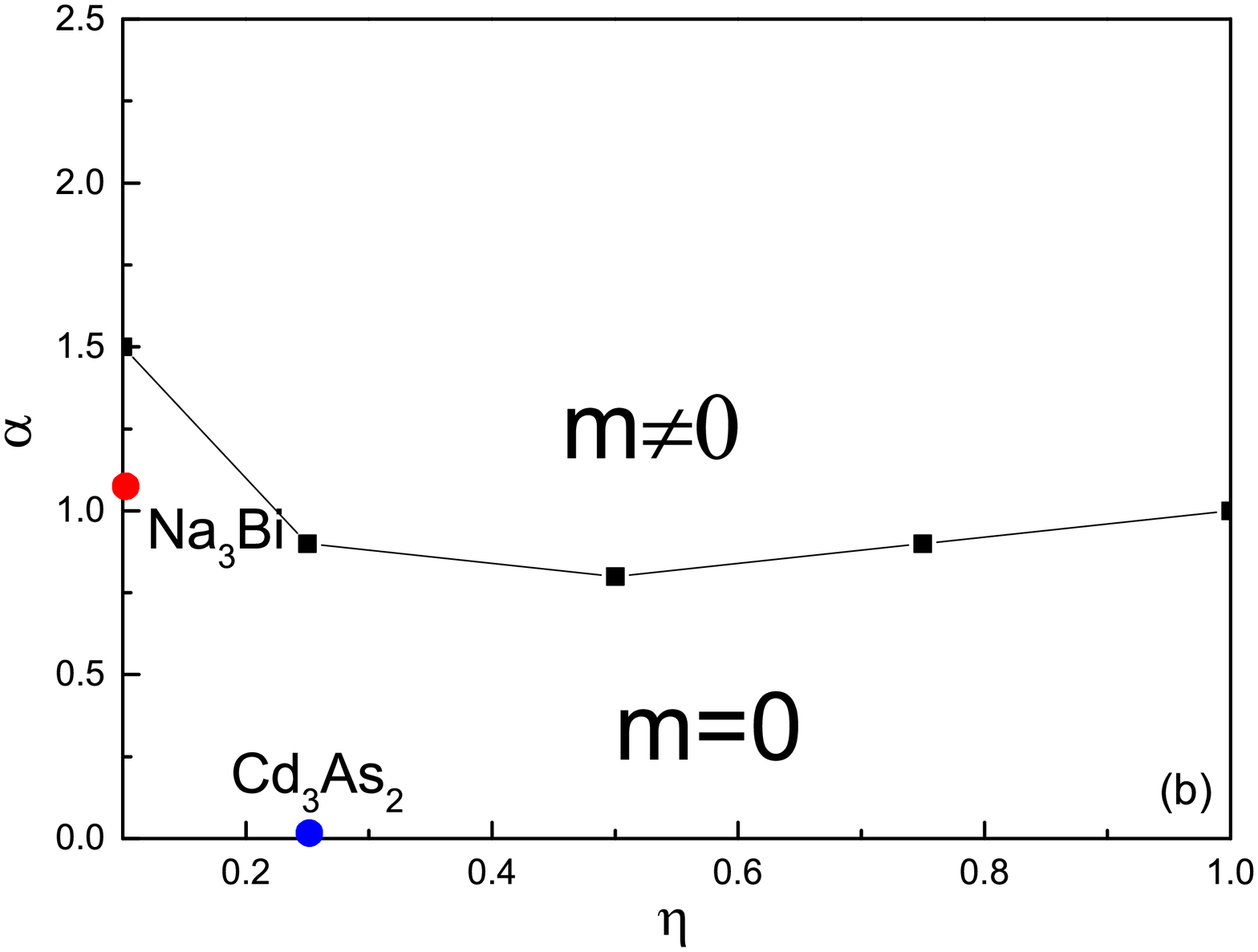}
\caption{Phase diagram on the $\alpha$-$\eta$ plane. (a)
Instantaneous approximation; (b) Khveshchenko approximation.}
\label{Fig:PhaseDiagramKhveshchenko}
\end{figure}

\begin{figure}
\includegraphics[width=0.36\textwidth]{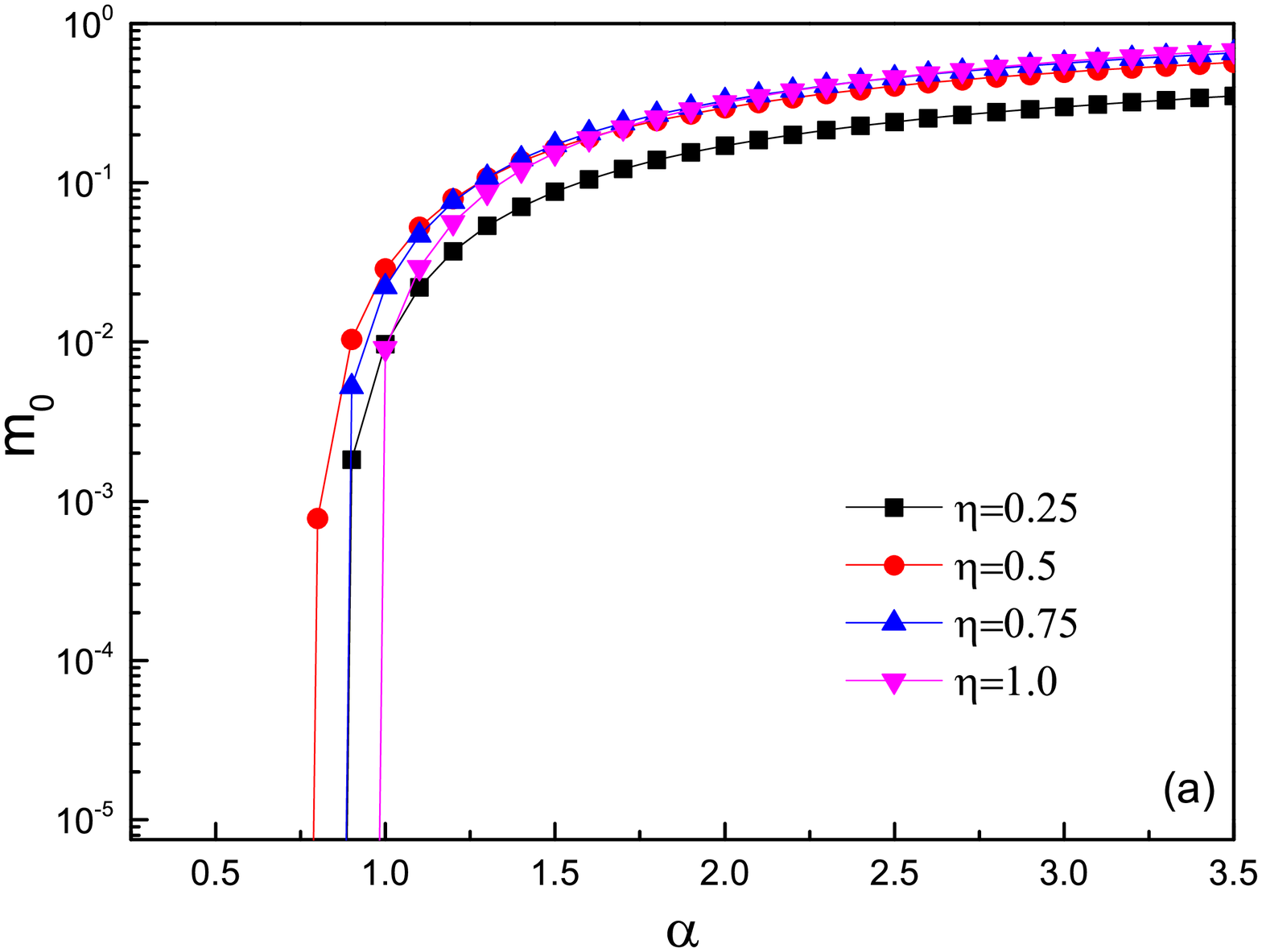}
\includegraphics[width=0.36\textwidth]{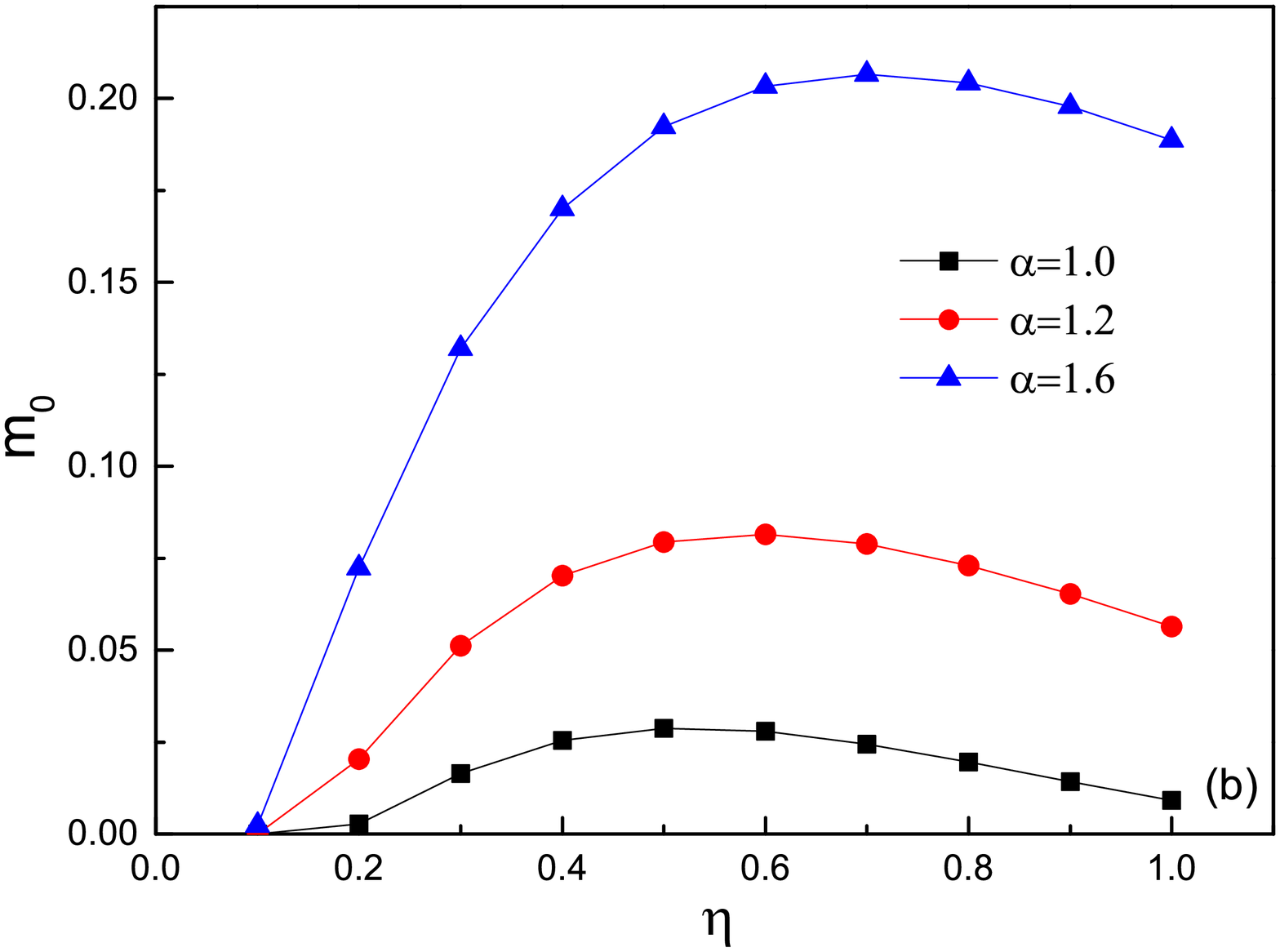}
\caption{(a) The $\alpha$-dependence of $m_{0}$ at different values
of $\eta$; (b) The $\eta$-dependence of $m_{0}$ at different values
of $\alpha$. Results are obtained under the Khveshchenko
approximation.}\label{Fig:Gap0Khveshchenko}
\end{figure}

We then numerically solve Eq.~(\ref{Eq:GapEquationKhveshchenko}) and
present the results in Fig.~\ref{Fig:Gap0Khveshchenko}. The
corresponding $\alpha$-$\eta$ phase diagram is given in
Fig.~\ref{Fig:PhaseDiagramKhveshchenko}(b). We observe that the
basic results are qualitatively the same as those obtained under the
instantaneous approximation. In particular, for any given value of
$\eta$, there is always a critical value $\alpha_c$ beyond which a
finite gap is generated, and the gap is a monotonously increasing
function of $\alpha$ in the range of $\alpha > \alpha_c$. For a
specific, sufficiently large $\alpha$, the gap exhibits a
non-monotonic dependence on the velocity ratio $\eta$, with its
maximum being reached at certain critical ratio $\eta_m$.

Although the conclusion is qualitatively the same, the quantitative
results obtained under the Khveshchenko approximation are different
from the instantaneous approximation. For instance, the critical
value $\alpha_{c} \approx 1.5$ for $\eta = 0.1$, and $\alpha_{c}
\approx 0.9$ for $\eta = 0.25$. In addition, $\alpha_{c} \approx
1.0$ for $\eta = 1$. The smallest value of $\alpha_{c}$ appears at
$\eta \approx 0.5$. Comparing
Fig.~\ref{Fig:PhaseDiagramKhveshchenko}(a) to
Fig.~\ref{Fig:PhaseDiagramKhveshchenko}(b), an apparent fact is that
$\alpha_c$ obtained under the Khveshchenko approximation is
generically slightly smaller than the one obtained under the
instantaneous approximation. Once again, we conclude that Na$_{3}$Bi
and Cd$_{3}$As$_{2}$ are both in the gapless semimetal phase.

\subsection{More suitable definitions of $\alpha$ and $\eta$}\label{Numerical:C}

\begin{figure}
\center
\includegraphics[width=0.363\textwidth]{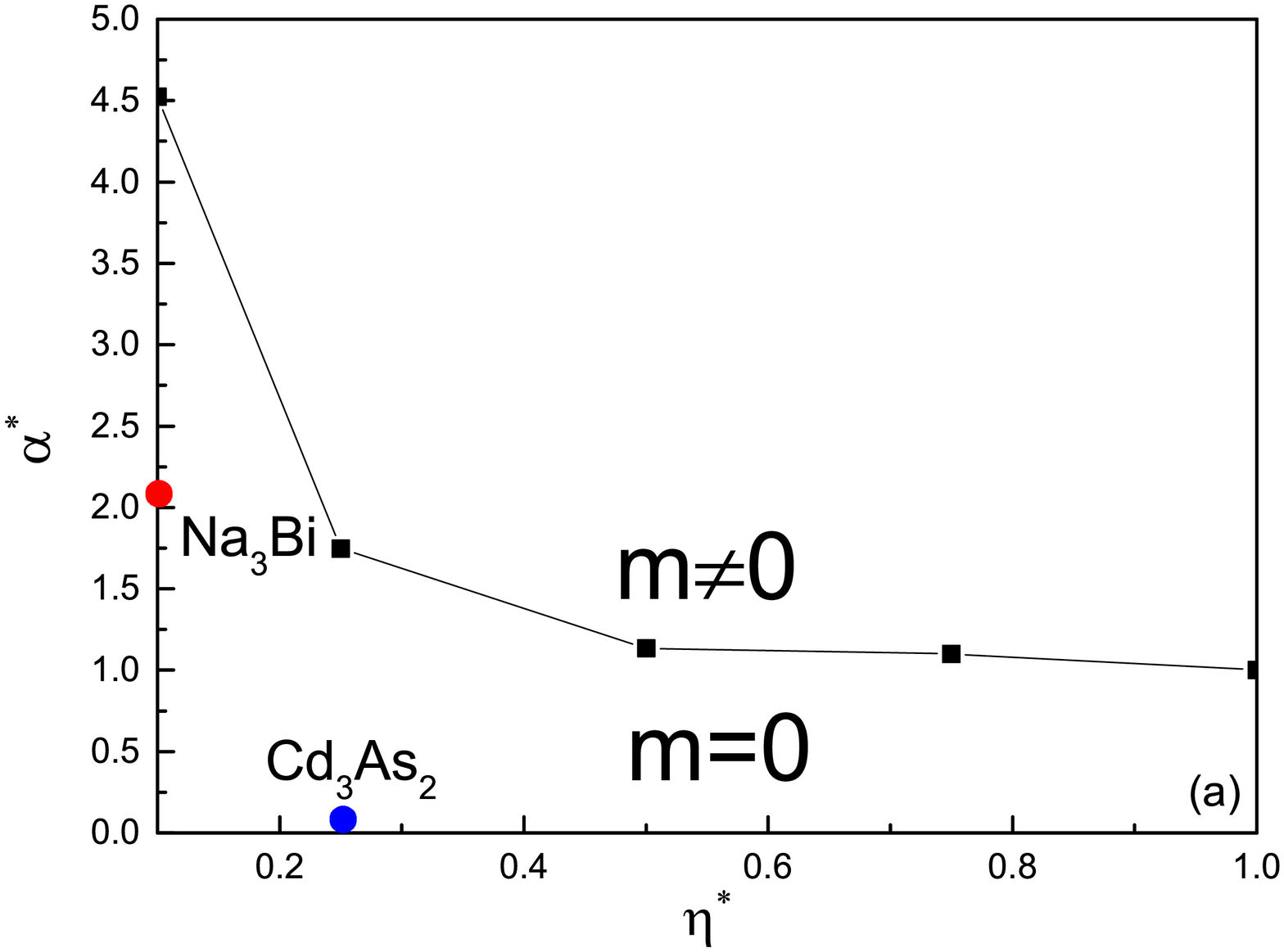}
\includegraphics[width=0.36\textwidth]{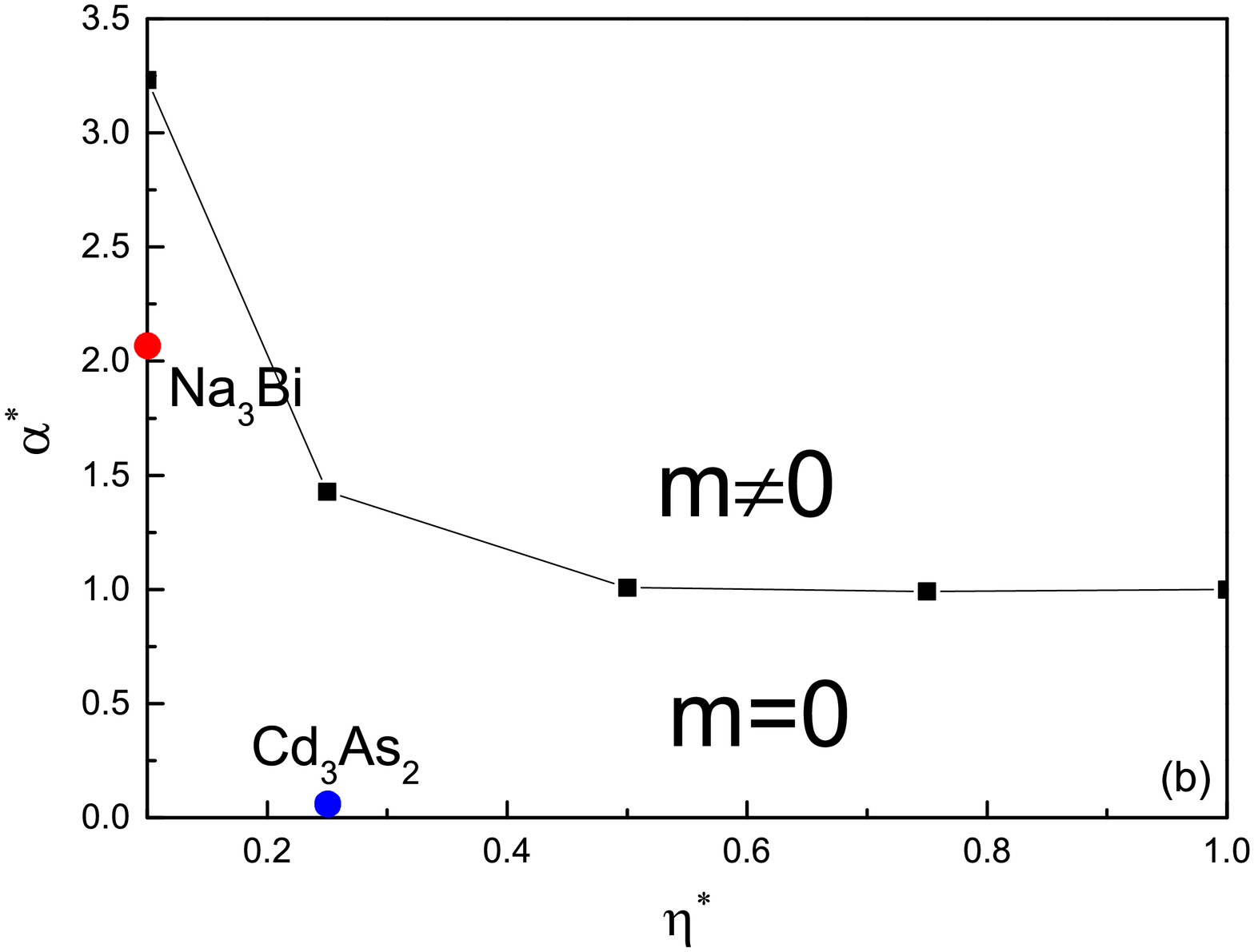}
\caption{Phase diagram on the $\alpha^{*}$-$\eta^{*}$ plane. (a)
Instantaneous approximation; (b) Khveshchenko approximation.}
\label{Fig:PhaseDiagramKhveshchenkoNew}
\end{figure}

In the above analysis, we have defined the interaction strength and
velocity ratio by $\alpha =
\frac{e^{2}}{v_{\parallel}\epsilon_{0}\epsilon_{r}}$ and $\eta =
\frac{v_{z}}{v_{\parallel}}$, respectively. These definitions were
introduced and utilized in previous works \cite{Braguta16,
Braguta17}. We would like to emphasize that these two definitions
might not be appropriate \cite{Xiao17}. For instance, to examine the
sole impact of the velocity anisotropy, one can fix the value of
$\alpha$, which means $v_{\parallel}$ is simultaneously fixed, and
tune the ratio $\eta$ by varying $v_{z}$. Because $v_{\parallel}$ is
fixed and $v_z$ is varying, the total kinetic energy of 3D Dirac
fermions are altered, and thus the effective strength of Coulomb
interaction, which is determined by the ratio between the potential
energy and the total kinetic energy, is also changed. Therefore, the
Coulomb interaction is automatically tuned by varying $\eta$, though
$\alpha$ remains fixed at a constant. As a consequence, the
influences of the Coulomb interaction and the velocity anisotropy
are entangled, and cannot be separated. In order to figure out how
the Coulomb interaction and the velocity anisotropy separately
affects dynamical gap generation, a more suitable choice is to
define
\begin{eqnarray}
\alpha^{*} = \frac{e^{2}}{\bar{v}\epsilon_{0}\epsilon_{r}} \quad
\mathrm{and} \quad \eta^{*} = \frac{v_{z}}{v_{\parallel}},
\end{eqnarray}
where $\bar{v} = \sqrt[3]{v_{\parallel}^{2}v_{z}}$ represents a mean
value of the fermion velocities. Now the two parameters $\alpha^{*}$
and $\eta^{*}$ can vary independently. Carrying out a simple
transformation of the results expressed by $\alpha$ and $\eta$, we
obtain a new phase diagram of 3D DSM depicted on the plane spanned
by $\alpha^{*}$ and $\eta^{*}$, as shown by
Fig.~\ref{Fig:PhaseDiagramKhveshchenkoNew}. We observe that, as the
velocity anisotropy increases, the critical interaction strength
grows dramatically. These results indicate that, the fermion
velocity anisotropy tends to suppress gap generation, and the
non-monotonic behavior shown in Fig.~(\ref{Fig:Gap0Instan})b and
Fig.~(\ref{Fig:Gap0Khveshchenko})b originates from the competition
between the increasing interaction strength and the growing velocity
anisotropy. The suppression of dynamical gap generation by
decreasing $\eta$ should be attributed to the enhanced dynamical
screening of Coulomb interaction.

\subsection{Impact of higher-order corrections\label{Sec:HighOrderCorrection}}

We have solved Eqs.~(\ref{Eq:DSEqRGV1})-(\ref{Eq:DSEqRGV3}) by
setting $\eta = 1$ and $N=2$. No dynamical gap is generated even
when $\alpha\rightarrow \infty$. It is important to notice that the
system contains two tuning parameters, namely $N$ and $\alpha$.
Excitonic pairing occurs only when $N<N_c$ and $\alpha > \alpha_c$.
If $N_c > 2$, one can simply fix $N=2$ and then determine $\alpha_c$
by solving DS equations. However, if $N_c < 2$, the Coulomb
interaction cannot trigger excitonic pairing even in the $\alpha
\rightarrow \infty$ limit. Actually, we find that $N_c \simeq 1.7$
in the limit $\alpha\rightarrow \infty$. It turns out that fermion
velocity renormalization tends to suppress dynamical gap generation.

\begin{figure}
\includegraphics[width=0.363\textwidth]{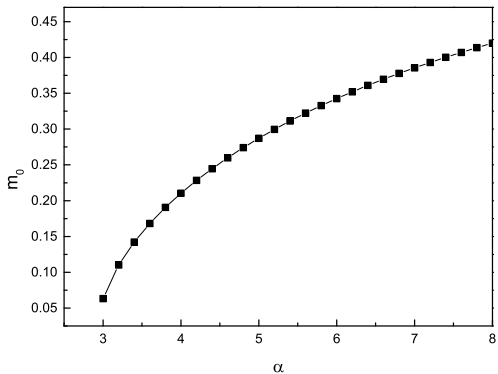}
\caption{The $\alpha$-dependence of $m_0$ obtained after solving
Eqs.~(\ref{Eq:DSESIsotropicA})-(\ref{Eq:DSESIsotropicC}) at $N=2$.
The critical value $\alpha_{c} \simeq 3.0$.}
\label{Fig:GapIsotropicWaveVeloRG}
\end{figure}

We emphasize here that the result $N_c < 2$ is obtained by ignoring
several potentially important effects, including the dynamical
screening of Coulomb interaction, the wave-function renormalization,
and the vertex correction, as evidenced by
Eq.~(\ref{Eq:VRGInstanCondition}). Such a result might be changed
considerably when these effects are taken into account. To determine
the influence of these corrections, we have solved
Eqs.~(\ref{Eq:DSESIsotropicA})-(\ref{Eq:DSESIsotropicC}) and find
that $N_{c}\simeq 4.2$. For physical flavor $N=2$, the dependence of
zero-energy gap $m_0$ on $\alpha$ is presented in
Fig.~\ref{Fig:GapIsotropicWaveVeloRG}, which clearly shows that
$\alpha_c \simeq 3.0$. For Na$_{3}$Bi and Cd$_{3}$As$_{2}$, the
fermion dispersion is strongly anisotropic and $\eta \ll 1$.
According to the results given in Sec.~\ref{Numerical:C}, the value
of $\alpha_c$ will be further increased as $\eta$ decreases from
$\eta=1$, which makes excitonic pairing more unlikely.

In order to calculate $\alpha_c$ and $N_c$ more accurately, it will
be necessary to incorporate even more corrections, such as the
feedback of fermion velocity renormalization and wave-function
renormalization on the polarization function. Incorporating all
these corrections is technically very involved, and will be studied
in a separate work. According to the extensive calculations carried
out by employing different approximations, it appears safe to
conclude that Na$_3$Bi and Cd$_3$As$_2$ are both deep in the
semimetallic phase, although more extensive calculations are needed
to precisely determine $\alpha_c$ and $N_c$.

\section{Summary and Discussion \label{Sec:Summary}}

In summary, we have studied the stability of the semimetal ground
state of 3D DSM against the long-range Coulomb interaction by making
a DS equation analysis. To the leading order of $1/N$ expansion, we
have solved the gap equation numerically and obtained a detailed
phase diagram on the plane spanned by the Coulomb interaction
strength and the velocity anisotropy parameter. Our results indicate
that, while excitonic gap generation is promoted as the interaction
becomes stronger, it is suppressed if the velocity anisotropy is
enhanced. As a concrete application of our results, we have
confirmed that the Coulomb interaction in Na$_{3}$Bi and
Cd$_{3}$As$_{2}$ is not strong enough to open a dynamical gap. Thus,
the semimetal ground state is very stable against Coulomb
interaction. In fact, these two 3D DSMs lie deep in the gapless
semimetal phase, hence the quantum fluctuation of excitonic pairing
is ignorable and does not lead to any detectable effect.

We also have examined the impact of several higher-order
corrections. In particular, we have incorporated the dynamical
screening of Coulomb interaction, the fermion velocity
renormalization, the wave-function renormalization, and the vertex
correction into the DS equations. The new critical value $\alpha_c$
is quantitatively different from that obtained by retaining only the
leading order of $1/N$ expansion. Nevertheless, the new $\alpha_c$
is still much larger than the physical value of $\alpha$ in
Na$_{3}$Bi and Cd$_{3}$As$_{2}$, implying that these two materials
are both robust gapless semimetals.

Recent Monte Carlo simulations \cite{Braguta16, Braguta17} reached
distinct conclusions concerning the strict ground state of
Na$_{3}$Bi and Cd$_{3}$As$_{2}$. A crucial difference between our
results and those obtained Ref.~\cite{Braguta16} and
Ref.~\cite{Braguta17} is in the chosen value of the dielectric
constant. The relative dielectric constant $\epsilon_r$ was
incorrectly missed in the calculations of Ref.~\cite{Braguta16} and
Ref.~\cite{Braguta17}. In fact, if the dielectric constant of
Na$_{3}$Bi and Cd$_{3}$As$_{2}$ are correctly chosen, the lattice
simulation result could be consistent with our conclusion and also
consistent with experiments.

It is interesting to search for the possible mechanism to promote
dynamical gap generation in realistic 3D DSM materials. Since
$\alpha \propto 1/(\bar{v}\epsilon_{r})$, the interaction will be
made stronger if one finds an efficient way to decrease $\bar{v}$
and/or $\epsilon_{r}$. For 2D materials, the value of $\epsilon_{r}$
is strongly affected by the substrate. For example, $\epsilon_{r}
\approx 2.8$ in graphene placed on SiO$_{2}$ substrate
\cite{CastroNetoPhysics09}, but $\epsilon_{r} = 1$ in suspended
graphene. However, this scenario does not work in 3D DSMs, because
changing the environment of a 3D material can hardly affect the
value of the bulk $\epsilon_{r}$. Recent theoretical study
\cite{Tang15} predicted that applying a uniform strain to graphene
might enhance the Coulomb strength by reducing the Dirac fermion
velocities. We speculate that this manipulation provides a promising
method to reinforce the Coulomb interaction of Na$_{3}$Bi and
Cd$_{3}$As$_{2}$. Another way to promote dynamical gap generation is
to find more 3D DSM materials other than Na$_{3}$Bi and
Cd$_{3}$As$_{2}$ that have smaller values of fermion velocities and
smaller $\epsilon_{r}$.

The Coulomb interaction strength is $\alpha \approx 0.06$ in
Cd$_{3}$As$_{2}$, which provides a small parameter to carry out
ordinary perturbative expansion. Previous perturbative calculations
\cite{Abrikosov70, Goswami11, Hosur12, Hofmann15, Throckmorton15}
revealed that the fermion velocity grows with lowering energy, and
that some observable quantities, including specific heat,
compressibility, optical conductivity, and susceptibility, exhibit
logarithmic-like dependence on energy or temperature. However, it is
important to emphasize that, the perturbative expansion method
cannot be used to compute the dynamical gap, because excitonic
pairing is a genuine non-perturbative phenomenon and should be
studied by means of non-perturbative tools, such as the DS equation
approach and the quantum Monte Carlo simulation \cite{Braguta16,
Braguta17, Karthik15}.

\acknowledgments

The authors acknowledge the financial support by the National
Natural Science Foundation of China under Grants No.11535005,
No.11475085, No.11690030, No.11504379, and No.11574285, and the
Fundamental Research Funds for the Central Universities under Grant
020414380074. G.-Z.L. is also supported by the Fundamental Research
Funds for the Central Universities (P. R. China) under Grant
WK2030040085.

\appendix

\begin{widetext}

\section{Calculation of the polarization \label{Sec:AppendixPola}}

We now provide the detailed calculation of the polarization function
that appears in the dressed Coulomb interaction function
Eq.~(\ref{Eq:CoulombExp}).

The free fermion propagator for massless Dirac fermion is given by
\begin{eqnarray}
G(\omega,\mathbf{k}) = \frac{1}{\omega\gamma_{0} +
v_{\parallel}(\gamma_{1}k_{x} + \gamma_{2}k_{y}) +
v_{z}\gamma_{3}k_{z}}. \label{Eq:FermionPropagatorConstantMass}
\end{eqnarray}
To the leading order of $1/N$ expansion, the polarization function
is defined as
\begin{eqnarray}
\Pi(\Omega,q_{x},q_{y},q_{z}) &=& N\int\frac{d\omega}{2\pi}
\frac{d^3\mathbf{k}}{(2\pi)^{3}}
\mathrm{Tr}\left[\gamma_{0}G(\Omega,\mathbf{k}) \gamma_{0}
G\left(i(\omega+\Omega),\mathbf{k}+\mathbf{q}\right)\right],
\label{Eq:PolaDefinition}
\end{eqnarray}
where $N$ is the fermion flavor. Substituting
Eq.~(\ref{Eq:FermionPropagatorConstantMass}) into
Eq.~(\ref{Eq:PolaDefinition}), we obtain
\begin{eqnarray}
\Pi\left(\Omega,\frac{q_{x}}{v_{\parallel}},\frac{q_{y}}{v_{\parallel}},
\frac{q_{z}}{v_{z}}\right) &=& \frac{4N}{v_{\parallel}^{2}v_{z}}\int
\frac{d\omega}{2\pi} \frac{d^3\mathbf{k}}{(2\pi)^{3}}
\frac{\omega(\omega+\Omega) - \mathbf{k}
\cdot(\mathbf{k}+\mathbf{q})}{\left(\omega^{2} +
\mathbf{k}^{2}\right)\left[(\omega+\Omega)^{2} +
\left|\mathbf{k}+\mathbf{q}\right|^{2}\right]},
\end{eqnarray}
where we have used the following transformations
\begin{eqnarray}
v_{\parallel}k_{x}\rightarrow k_{x},\quad
v_{\parallel}k_{y}\rightarrow k_{y}, \quad
v_{z}k_{z}\rightarrow k_{z}, \quad
v_{\parallel}q_{x}\rightarrow q_{x}, \quad
v_{\parallel}q_{y}\rightarrow q_{y},\quad v_{z}q_{z}\rightarrow
q_{z}.
\end{eqnarray}
Making use of the Feynman parametrization formula
\begin{equation}
\frac{1}{AB} = \int_{0}^{1}dx\frac{1}{\left[xA+(1-x)B\right]^2},
\end{equation}
we get
\begin{eqnarray}
\Pi\left(\Omega,\frac{q_{x}}{v_{\parallel}},
\frac{q_{y}}{v_{\parallel}},\frac{q_{z}}{v_{z}}\right) &=&
\frac{4N}{v_{\parallel}^{2}v_{z}}\int_{0}^{1}dx \int
\frac{d\omega}{2\pi} \frac{d^3\mathbf{k}}{(2\pi)^{3}}
\frac{\omega(\omega + \Omega)-\mathbf{k}\cdot(\mathbf{k} +
\mathbf{q})}{\left[\left(\omega+x\Omega\right)^2 + |\mathbf{k} +
x\mathbf{q}|^{2}+x(1-x)\left(\Omega^{2} + \mathbf{q}^{2}
\right)\right]^{2}}.
\end{eqnarray}
We then re-define $\omega'=\omega+x\Omega$ and $\mathbf{k}' =
\mathbf{k} + x\mathbf{q}$, and re-write the polarization in the form
\begin{eqnarray}
\Pi\left(\Omega,\frac{q_{x}}{v_{\parallel}},
\frac{q_{y}}{v_{\parallel}},\frac{q_{z}}{v_{z}}\right) &=&
\frac{4N}{v_{\parallel}^{2}v_{z}}\int_{0}^{1}dx \int
\frac{d\omega'}{2\pi}\frac{d^3\mathbf{k}'}{(2\pi)^{3}}
\frac{\omega'^2-\mathbf{k}'^{2} - x(1-x)\left(\Omega^{2} -
\mathbf{q}^{2}\right) }{\left[\omega'^2+\mathbf{k}'^{2} +
x(1-x)\left(\Omega^{2}+\mathbf{q}^{2}\right)\right]^{2}}.
\end{eqnarray}
After carrying out the integration over $\omega'$ and momenta, we
get
\begin{eqnarray}
\Pi\left(\Omega,\frac{q_{x}}{v_{\parallel}},
\frac{q_{y}}{v_{\parallel}},\frac{q_{z}}{v_{z}}\right) &=& \frac{2N
\mathbf{q}^{2}}{v_{\parallel}^{2}v_{z}}\int_{0}^{1} dx x(1-x) \int
\frac{d^3\mathbf{k}'}{(2\pi)^{3}} \frac{1}{\left[\mathbf{k}'^{2} +
x(1-x)\left(\Omega^{2}+\mathbf{q}^{2}\right)\right]^{3/2}}
\nonumber \\
&=& \frac{2N \mathbf{q}^{2}}{\pi^2 v_{\parallel}^{2}v_{z}}F,
\end{eqnarray}
where
\begin{eqnarray}
F &=& \left\{\frac{1}{12}\ln\left(\frac{2\Lambda+\sqrt{4\Lambda^{2}
+ \left(\Omega^{2}+\mathbf{q}^{2}\right)}}{\sqrt{\Omega^{2} +
\mathbf{q}^{2}}}\right) - \int_{0}^{\frac{1}{2}}dx x(1-x)
\frac{\Lambda}{\sqrt{\Lambda^{2} + x(1-x)\left(\Omega^{2} +
\mathbf{q}^{2}\right)}}\right.
\nonumber \\
&&-\frac{1}{12}\int_{0}^{\frac{1}{2}}dx\left(3x^2-2x^3\right)
\frac{(1-2x)\left(\Omega^2+\mathbf{q}^{2}\right)}{\left[\Lambda +
\sqrt{\Lambda^{2}+x(1-x)\left(\Omega^{2}+\mathbf{q}^{2}\right)}\right]
\sqrt{\Lambda^{2}+x(1-x)\left(\Omega^{2}+\mathbf{q}^{2}\right)}}\nonumber
\\
&&\left.+\frac{1}{12}\int_{0}^{\frac{1}{2}}dx\left(3x^2-2x^3\right)
\frac{(1-2x) \left(\Omega^2+\mathbf{q}^{2}\right)}{x(1-x)
\left(\Omega^{2}+\mathbf{q}^{2}\right)} \right\}.
\end{eqnarray}
In the regime $\sqrt{\Omega^{2}+\mathbf{q}^{2}}\ll \Lambda$, we
retain only the leading term, i.e.,
\begin{eqnarray}
\Pi\left(\Omega,\frac{q_{x}}{v_{\parallel}},\frac{q_{y}}{v_{\parallel}},
\frac{q_{z}}{v_{z}}\right) = \frac{N \mathbf{q}^{2}}{6\pi^2
v_{\parallel}^{2}v_{z}}\ln\left(\frac{\Lambda}{\sqrt{\Omega^{2} +
\mathbf{q}^{2}}}\right).
\end{eqnarray}
Introducing the re-definitions $q_{x}\rightarrow
v_{\parallel}q_{x}$, $q_{y}\rightarrow
v_{\parallel}q_{y}$, $q_{z}\rightarrow v_{z}q_{z}$, and
$\Lambda\rightarrow
\left(v_{\parallel}^{2}v_{z}\right)^{1/3}\Lambda$, we have
\begin{eqnarray}
\Pi(\Omega,q_{\parallel},q_{z}) =
\frac{N\left(v_{\parallel}^{2}q_{\parallel}^{2} +
v_{z}^{2}q_{z}^{2}\right)}{6\pi^2v_{\parallel}^{2}v_{z}}
\ln\left(\frac{\left(v_{\parallel}^{2}v_{z}\right)^{1/3}
\Lambda}{\sqrt{\Omega^{2}+v_{\parallel}^{2}q_{\parallel}^{2}
+ v_{z}^{2}q_{z}^{2}}}\right). \label{Eq:PolaExpA}
\end{eqnarray}
Dynamical gap generation is a low-energy phenomenon, and the
dominant contribution to the gap equation comes from the small
enegy/momenta regime. Although the contribution from high
energy/momenta regime is unimportant, the approximate polarization
should be at least well-defined. We notice that the above
approximate expression of $\Pi(\Omega,q_{\parallel},q_{z})$ is
negative at very high energies, i.e., $\Omega \gg
\left(v_{\parallel}^{2}v_{z}\right)^{1/3}\Lambda$, which would lead
to a unphysical pole in the dressed Coulomb interaction function.
The exact polarization is definitely always positive. Such
unphysical pole originates from an improper approximation. In order
to avoid the appearance of such pole, we make the following
replacement
\begin{eqnarray}
\ln\left(\frac{\left(v_{\parallel}^{2}v_{z}\right)^{1/3}
\Lambda}{\sqrt{\Omega^{2}+v_{\parallel}^{2}q_{\parallel}^{2}
+ v_{z}^{2}q_{z}^{2}}}\right)\rightarrow
\ln\left(\frac{\left(v_{\parallel}^{2}v_{z}\right)^{1/3}
\Lambda+ \sqrt{\Omega^{2}+v_{\parallel}^{2}
q_{\parallel}^{2} + v_{z}^{2} q_{z}^{2}}}{\sqrt{\Omega^{2} +
v_{\parallel}^{2}q_{\parallel}^{2} + v_{z}^{2}
q_{z}^{2}}}\right).
\end{eqnarray}
Now the polarization becomes
\begin{eqnarray}
\Pi(\Omega,q_{\parallel},q_{z}) =
\frac{N\left(v_{\parallel}^{2}q_{\parallel}^{2} +
v_{z}^{2}q_{z}^{2}\right)}{6\pi^2v_{\parallel}^{2}v_{z}}
\ln\left(\frac{\left(v_{\parallel}^{2}v_{z}\right)^{1/3}
\Lambda +
\sqrt{\Omega^{2}+v_{\parallel}^{2}q_{\parallel}^{2} +
v_{z}^{2} q_{z}^{2}}}{\sqrt{\Omega^{2} + v_{\parallel}^{2}
q_{\parallel}^{2} + v_{z}^{2}q_{z}^{2}}}\right).
\label{Eq:PolaExpB}
\end{eqnarray}
This new polarization is very close to the exact polarization in the
low energy/momenta regime, and meanwhile does not yield any
unphysical pole in the high energy/momenta regime. We have used this
approximate polarization in our DS equation calculations.
\end{widetext}

\end{document}